\RequirePackage{ifpdf}
\documentclass[12pt,letterpaper]{JHEP3}
\pdfoutput=1
\usepackage{cite}
\usepackage{epsfig}
\usepackage{graphicx}
\usepackage{subfigure}

\graphicspath{{Figures/}}

\usepackage{color}
\newcommand{\Beq}{\begin{eqnarray}}
\newcommand{\Eeq}{\end{eqnarray}}

\newcommand{\sech}{\mathrm{sech}}

\title{Ultra-relativistic oscillon collisions}

\author{Mustafa A. Amin${}^{1}$\footnote{mustafa.a.amin@gmail.com}, Indranil Banik${}^2$, Carina Negreanu${}^3$ and 
I-Sheng Yang${}^4$\footnote{isheng.yang@gmail.com} \\
${}^{1,2,3}$ Kavli Institute for Cosmology and Institute of Astronomy, Madingley Rd, Cambridge CB3 0HA, United Kingdom\\
${}^{4}$IOP and GRAPPA, Universiteit van Amsterdam, Science Park 904, 1090 GL Amsterdam, Netherlands
}

\abstract
{
In this short note we investigate the ultra-relativistic collisions of small amplitude oscillons in 1+1 dimensions. Using the amplitude of the oscillons and the inverse relativistic boost factor $\gamma^{-1}$ as the perturbation variables, we analytically calculate the leading order spatial and temporal phase shifts, and the change in the amplitude of the oscillons after the collisions. At leading order, we find that only the temporal phase shift receives a nonzero contribution, and that the collision is elastic. This work is also the first application of the general kinematic framework for understanding ultra-relativistic collisions \cite{ALY13} to intrinsically time-dependent solitons.
} 
\begin{document}
\section{Introduction}

The dynamics of soliton collisions can be complex because of the necessary nonlinearity of the equations governing them. Some of us recently demonstrated that in certain nonlinear scalar field theories, an analytical formalism is possible for understanding ultra-relativistic soliton collisions \cite{ALY13}. The accuracy of this formalism was demonstrated for colliding (anti)kinks in periodic potentials \cite{ALY13a}. The examples in those papers emphasized the generality of the formalism with respect to general periodic potentials (arbitrarily far away from known integrable ones). Although these were limited to $(1+1)$ dimensional stationary solitons (ie. intrinsically time-independent solutions), neither the dimensionality nor the stationarity is a fundamental limitation of the formalism. Here, we demonstrate the applicability of this formalism to intrinsically time-dependent solitons, which in addition do not rely on the periodicity of the potential. As a concrete example, we study the ultra-relativistic collision of oscillons: spatially localized, oscillatory in time and unusually long-lived solutions of the nonlinear Klein-Gordon equation (for example, see \cite{Bogolyubsky:1976yu,Gleiser:1993pt,CopGle95}).
Generalization to higher dimensions is also an interesting direction\footnote{This includes, for example, the right-angle vortex scattering that is relevant for string intercommutation\cite{She87,JacJon04,CopKib06,AchVer10}.}, but we will leave that for future work.

The importance of this new example is twofold. First of all, due to Derrick's theorem \cite{Derrick:1964}, many simple field theories cannot support stationary solitons in three dimensions or higher. Localized, time-dependent solitons are the most general objects that the kinematic scattering framework in \cite{ALY13,ALY13a} applies to. Secondly, these particular solitons (oscillons) appear in a wide range of physical scenarios. They can be produced copiously at the end of inflation, in bubble collisions and in phase transitions in the early universe \cite{Hawking:1982ga,AguJoh08,JohYan10,Gleiser:2011xj,Amin:2011hj}. They can delay thermalization, play a role in baryogenesis \cite{Lozanov:2014zfa}, might appear in dark-matter/dark energy models \cite{AmiZuk11} and are also found in condensed matter systems (for example \cite{Aubry1997201}).  Given their wide-ranging applications, it is worthwhile to understand their interactions.

Oscillon collisions have been studied before (for example, see \cite{Campbell198647,Hindmarsh:2006ur,Hindmarsh:2007jb}). Controlled analytic calculations, however, have not been provided (to the best of our knowledge). Here we take a step towards an analytic understanding of their interactions in the small amplitude and ultra-relativistic limit in 1+1 dimensions. We will use the small amplitudes as well as the inverse relativistic boost factor $\gamma^{-1}$ as small perturbation variables to aid our calculations. Although ultra-relativistic, small-amplitude, 1+1 dimensional oscillon collisions are not typical in the physical scenarios discussed earlier, we hope that our formalism and results will lead to a better understanding of the general interaction dynamics.

We emphasize that our scenario has one important physical difference compared to earlier applications \cite{ALY13a}. When the background object is a stationary soliton, perturbations around the soliton (generated by the collision) admit a well-defined expansion in terms of separable eigenmodes. Here the oscillon background depends on both space and time, so the equation of motion for small perturbations is generically non-separable. That means there is no natural eigenbasis of perturbation modes. However, even in our intrinsically time-dependent situation, there are three modes which have a clear physical interpretation. The first two are the zero modes corresponding to the space and time translational symmetries of the oscillon. The third one is a small change in its amplitude, which is always possible since oscillons exist for a continuous range of small amplitudes. We will calculate the following leading order results for a stationary oscillon with an amplitude $\epsilon\ll 1$, temporal oscillation frequency $\omega =\sqrt{1-\epsilon^2}$ and spatial width $\sim \epsilon^{-1}$ that undergoes a collision with an incoming, ultra-relativistic oscillon (with $\gamma^{-1}=\sqrt{1-v^2}\ll 1$) and amplitude $\epsilon_i\ll 1$:

\begin{itemize}
\item the change of internal oscillation phase, $\omega\Delta t =4\epsilon_i/\gamma$,
\item the shift in position compared to the oscillon width, $\epsilon \Delta x=0$,
\item the relative change in amplitude, $\Delta\epsilon/\epsilon =0$.
\end{itemize}
Note that the second point implies no velocity change (no time dependence in the position shift), and the last point means no change in the internal energy of the oscillation. Together, these two indicate that such collisions are elastic at leading order.
\section{Small amplitude oscillons}
\label{sec:oscillons}
Oscillons are time-dependent, localized, (pseudo-)solitonic configurations that are found in many scalar field theories with nonlinear couplings. In this paper we will focus on a simple and well-studied model in $(1+1)$ dimensions:
\Beq
\mathcal{L} &=& -\frac{1}{2}(\partial_t\phi)^2+\frac{1}{2}(\partial_x\phi)^2 +V(\phi),\\
V(\phi)&=&\frac{1}{2}\phi^2-\frac{1}{4}\phi^4+... \label{eq-genL}
\Eeq
where we have assumed a symmetric potential\footnote{We started with the potential $$\mathcal{L}=-\frac{1}{2}(\partial_t\phi)^2 + \frac{1}{2}(\partial_x\phi)^2 + \frac{m^2}{2}\phi^2-\frac{\lambda_4}{4} \phi^4.$$ We then expressed spatial lengths  and time intervals in units units of the mass $m$, and rescaled the fields by $m/\sqrt{\lambda_4}$.}. The minus sign in front of the quartic term (the {\it opening up} of the potential) is necessary for spatially localized solutions to exist. The equation of motion is
\Beq
(\partial_x^2-\partial_t^2)\phi=V'(\phi).
\Eeq
For the above equation, a  long-lived, localized and oscillatory solution (an oscillon) is given by (see for example: \cite{Amin:2013ika})
\Beq
\phi(x,t;\epsilon) &=& \epsilon 
\sqrt{\frac{8}{3}}~ 
{\rm sech}(\epsilon x) \cos\left(\omega t\right)
+ \mathcal{O}(\epsilon^3)~,\\
\omega&=&\sqrt{1-\epsilon^2},
\Eeq
where a single, small parameter $\epsilon\ll 1$ determines the amplitude, frequency ($\omega$) and the spatial width of this oscillon ($\sim \epsilon^{-1}$).
\subsection{Perturbed oscillon and zero modes}
We now consider small perturbations $h(x,t)$ around this oscillon solution. We assume that $h$ is small compared to the leading order term in the oscillon profile, but large compared to the higher order terms: $\epsilon^3\ll h \ll \epsilon$. The perturbation $h$ satisfies 
\begin{equation}
(\partial_x^2-\partial_t^2) h = 
[V'(\phi + h)-V'(\phi)]
\approx V''(\phi)h \approx [1+\mathcal{O}(\epsilon^2)]h~.
\label{eq-hmode}
\end{equation}
Within this function $h(x,t)$, two parameters ($\Delta x$ and $\Delta t$) quantify the spatial and temporal shifts of the oscillon solution:
\Beq
\phi(x-\Delta x,t-\Delta t;\epsilon) 
- \phi(x,t;\epsilon)\approx \epsilon \sqrt{\frac{8}{3}}\sech(\epsilon x)\left[(\omega\Delta t)\sin\left(\omega t\right)+(\epsilon\Delta x)\frac{\sinh(\epsilon x)}{\cosh(\epsilon x)}\cos\left(\omega t\right)\right].\nonumber
\Eeq  
At the leading order of $\epsilon$, these changes can be represented by two separable mode functions:
\begin{equation}
h(x,t)=g_1(x)f_1(t) + g_2(x)f_2(t) 
+ ...~,
\end{equation}
where 
\begin{eqnarray}
g_1(x) &=& \sqrt{\frac{\epsilon}{2}}{\rm sech}(\epsilon x)~, 
\label{eq-tmode} \\
g_2(x) &=& \sqrt{\frac{3\epsilon}{2}}
\frac{\sinh(\epsilon x)}{\cosh^2(\epsilon x)}~,
\label{eq-xmode}
\end{eqnarray}
and
\begin{eqnarray}
f_1(t) &=& \left(\omega\Delta t\right)4\sqrt{\frac{\epsilon}{3}}\sin\left(\omega t\right),
\label{eq-ft} \\ 
f_2(t) &=& (\epsilon\Delta x)\frac{4}{3}\sqrt{\epsilon}
\cos\left(\omega t\right)~.
\label{eq-fx}
\end{eqnarray}
Note that the particular normalizations ensure $\int g_n^2(x)dx =1$ for $n=1,2$. We can project any small perturbations onto these two modes and evaluate $\Delta x$ and $\Delta t$. Mode functions representing additional changes, which by definition are orthogonal to the spatial and temporal translations, are not needed for calculating the position and time shifts due to the collision.

For how a collision leads to the changes in the field profile, we do not need to include a change in amplitude $\Delta \epsilon$ explicitly in the leading order calculation. This is due to the energy-conservation/optical-theorem of the formalism in \cite{ALY13}. At the leading order of $\gamma^{-1}$, the time dependence in the position shift, $\Delta \dot{x}$, and the amplitude change $\Delta \epsilon$, are the only two contributions to the energy change. Other energy changes are ``leaks'', which can only appear at the second order or higher. Therefore, if we explicitly calculate $\Delta x$, we can then infer $\Delta \epsilon$ from it. In our particular case, as we will see, $\Delta x=0$, which directly means that $\Delta \epsilon=0$.

\begin{figure}[t!] 
   \centering
   \includegraphics[width=5in]{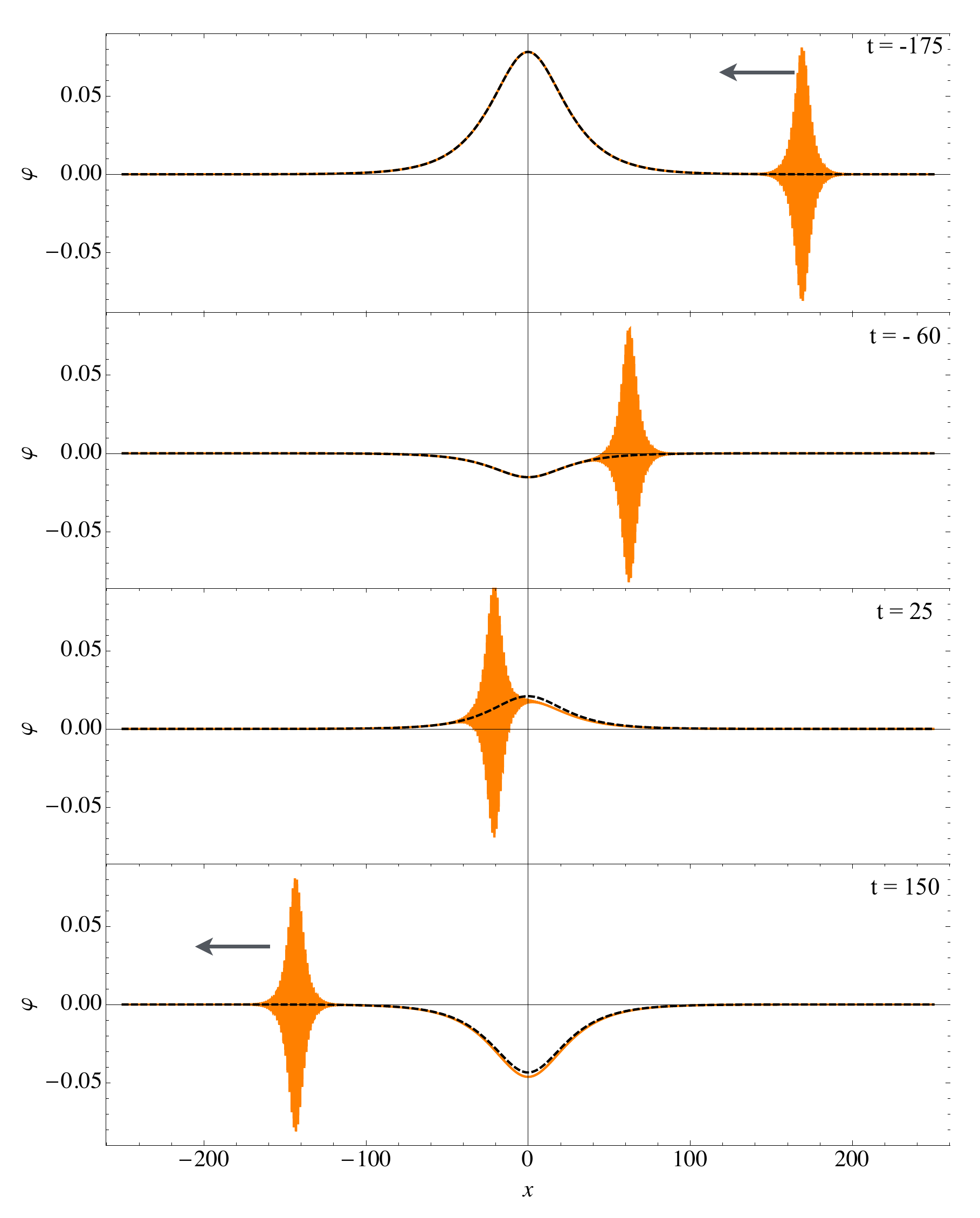} 
   \caption{The above figure shows an ultra-relativistic collision between two small amplitude oscillons. The incoming oscillon shows a large number of spatial oscillations due the Lorentz transformation of the oscillatory (temporal) part of the oscillon. The relative narrowness of the incoming oscillon is due to the Lorentz contraction of the spatial profile. The dashed line indicates the unperturbed solution for the stationary oscillon. Note that there is a distinct shift in the temporal phase of the stationary oscillon after the collision. This shift in temporal phase can be seen as the difference between the orange and black-dashed profiles of the stationary oscillons after the collision.}
   \label{fig:Collision}
\end{figure}
\section{Oscillon Collisions}
\label{sec:collision}
Consider a stationary oscillon $\phi_0$ with an amplitude $\epsilon$ centered around the origin. At early times, there is an incoming oscillon $\phi_i$ with amplitude $\epsilon_i$ moving towards the stationary one with a speed $v$ and boost factor $\gamma =1/\sqrt{1-v^2}$. Since field profile decays exponentially at the length scale of $\epsilon^{-1}$, we can simply add the profiles of oscillons separated by a distance $\gg\epsilon^{-1}$.  Hence, when the two oscillons are still far apart before the collision, we have
\begin{equation}
\phi_0 + \phi_i
= \phi(x,t;\epsilon) + 
\phi\left[\gamma(x+vt),\gamma(t+vx);\epsilon_i\right]~.
\end{equation}
These solitons collide with each other at around $t=0$. We capture four snapshots from a numerical simulation of this collision process in Fig.\ref{fig:Collision}. The changes in the stationary soliton, in particular a temporal phase shift, due to the collision is also clearly visible. We turn to the calculation of these collision related changes next.

\subsection{Collision as a perturbative theory of modes}

In the presence of an incoming oscillon, Eq.~(\ref{eq-hmode})  is modified as follows:
\begin{eqnarray}
\label{eq-heom}
(\partial_x^2-\partial_t^2 - 1)h 
&=& V'(\phi_0+\phi_i)-V'(\phi_0)-V'(\phi_i)~,\\ \nonumber
&=& -3(\phi_0^2\phi_i+\phi_0\phi_i^2)~,\\ \nonumber
&\equiv& S(x,t)~.
\end{eqnarray}
This equation is  accurate at the ``leading order'' in three small parameters, $\epsilon$, $\epsilon_i$ and $1/\gamma$. In the full equation of motion for $h$, the incoming oscillon also modifies the LHS. However, as argued in \cite{ALY13}, since $h$ started to be zero, it must first be sourced before any self-coupling becomes important. So modifications to the LHS  can only affect the result at higher order in the small parameters. 

Solving Eq.~(\ref{eq-heom}) is not significantly different from solving the full problem, since it still involves solving a PDE in two variables $x$ and $t$. The useful insight from \cite{ALY13} is that to calculate changes such as phase shifts, we can first ``project'' this equation of motion using the relevant spatial mode functions and then solve the corresponding ODE for the time-dependent amplitude of these spatial mode functions. The projection onto the $g_1$ mode yields
\begin{eqnarray}
& &\int dx g_1(x)(\partial_x^2-\partial_t^2 - 1)h(x,t)=\int dx g_1(x)S(x,t)~, \nonumber \\
\Longrightarrow & &-(\partial_t^2+1)f_1(t)+\int dx \partial_x^2g_1(x)h(x,t)= S_1(t)~,
\nonumber \\ \nonumber\\
\Longrightarrow& & -(\partial_t^2+1)f_1(t) + \mathcal{O}(\epsilon^2)=S_1(t)~,
\label{eq-teom}
\end{eqnarray}
where
\Beq
S_1(t)\equiv \int dx g_1(x)S(x,t)~.
\Eeq
Note that we have integrated by parts twice, and $\partial_x^2$ acting on the slowly varying profile $g_1$ suppresses it by a factor of $\epsilon^2$. The remaining leading order equation becomes a simple ODE for $f_1$. Similarly, we can project onto $g_2(x)$ to get
\Beq
-(\partial_t^2+1)f_2(t) + \mathcal{O}(\epsilon^2)=S_2(t)~,
\label{eq-xeom}
\Eeq
where
\Beq
S_2(t)\equiv \int dx g_2(x)S(x,t)~.
\Eeq
Solving Eq.~(\ref{eq-teom}) and (\ref{eq-xeom}) will allow us to calculate $\Delta x$ and $\Delta t$ created by the collision.

\subsection{Mode function solutions}

It is straightforward to write the solution of Eq.~(\ref{eq-teom}) and (\ref{eq-xeom}) as
\Beq
f_n(t)=-\int_{\infty}^t d\tau \sin (t-\tau)S_n(\tau).
\Eeq
where $n=1,2$. We can calculate the $S_n$ (and hence $f_n$) explicitly if we assume $\epsilon_i\gamma\gg\epsilon$. This condition is satisfied when  the two oscillons are about the same size and the collision is fast. The condition is slightly more general than that; we require that the length contracted profile of the incoming oscillon is much narrower than the stationary one. With these assumptions:
\begin{eqnarray}
S_1(t) &=& \label{eq-Stime}
-\int dx~3(\phi_0\phi_i^2+\phi_0^2\phi_i)g_1(x)~, \nonumber\\
&\approx&-\int dx~3\phi_0\phi_i^2g_1(x)~, 
\\ \nonumber 
&\approx& -\frac{3}{2}\epsilon\sqrt{\frac{8}{3}}
\sech(-\epsilon vt)\cos\left(\sqrt{1-\epsilon^2}t\right)g_1(-vt) 
\int dx~\epsilon_i^2\frac{8}{3}
\sech^2[\epsilon_i \gamma(x+vt)]~. \\ \nonumber
&=& -16\frac{\epsilon\epsilon_i}{\gamma}
\sqrt{\frac{\epsilon}{3}}
\sech^2(-\epsilon vt)\cos\left(\sqrt{1-\epsilon^2}t\right)~.
\end{eqnarray}

Note that a highly boosted oscillon, $\phi_i$, will have a rapidly oscillatory profile in $x$ with a spatial oscillation frequency $\propto \gamma$. As a result the term linear in $\phi_i$ does not contribute to the integral in the second line. $\phi_i^2$ on the other hand, is positive definite and will contribute. Nevertheless, we can treat $\phi_i^2$ as an averaged envelope instead of a rapidly oscillating profile, which is the approximation used in the third line. Due to the highly Lorentz contracted extent of $\phi_i^2$ centered around $x=-vt$, we treat other factors which are varying much more slowly in space as a time dependent height (with $x\rightarrow -vt$) of this averaged $\phi_i^2$ envelope. In the fourth line we have carried out the $x$ integral explicitly and use the normalized mode function $g_1(x)$ (see Eq.~(\ref{eq-tmode})). 

A similar calculation for the source term $S_2(t)$ in Eq. (\ref{eq-xeom}) yields
\begin{eqnarray}
S_2(t)&=&
-\int dx~3(\phi_0\phi_i^2+\phi_0^2\phi_i)g_2(x)~,\nonumber \\
&\approx& -16\frac{\epsilon\epsilon_i\sqrt{\epsilon}}{\gamma}~
\cos\left(\sqrt{1-\epsilon^2}t\right)
\frac{\sinh(-\epsilon v t)}{\cosh^3(-\epsilon v t)}~.
\end{eqnarray}

We can now solve for the $f_n$'s explicitly:
\begin{eqnarray}
f_1(t) &=& \nonumber
-\int_{-\infty}^t \sin(t-\tau)S_1(\tau)d\tau~, \\ \nonumber
&=& 16\frac{\epsilon\epsilon_i\sqrt{\epsilon}}
{\gamma\sqrt{3}}~
\int_{-\infty}^t\sin(t-\tau) \sech^2(-\epsilon v\tau)
\cos\left(\sqrt{1-\epsilon^2}\tau\right) d\tau~, \\ \nonumber
&=& 16\frac{\epsilon\epsilon_i\sqrt{\epsilon}}
{\gamma\sqrt{3}}~
\sin(t)\int_{-\infty}^\infty \cos(\tau)
\sech^2(-\epsilon v\tau)
\cos\left(\sqrt{1-\epsilon^2}\tau\right) d\tau~,
\nonumber \\
&=& 16\frac{\epsilon\epsilon_i\sqrt{\epsilon}}
{\gamma\sqrt{3}}~
\sin(t)\int_{-\infty}^\infty \frac{\sech^2(-\epsilon v\tau)}{2} d\tau~,\nonumber\\
&=&16\frac{\epsilon_i\sqrt{\epsilon}}
{\gamma\sqrt{3}}~\sin(t)~.
\label{eq-tshift}
\end{eqnarray}
In the third line we are assuming that $t\gg \epsilon^{-1}$, i.e. there is no significant overlap between the solitons and the collision is complete. In the last line, we combined two cosines, since they have the same frequency (to the leading order in $\epsilon$) and oscillate much faster than the slowly varying `sech' envelope. An almost identical calculation for $f_2(t)$ yields
\begin{eqnarray}
f_2(t) &=& 
-\int_{-\infty}^t \sin(t-\tau)S_2(\tau)d\tau~, \nonumber\\ 
&=& 16\frac{\epsilon\epsilon_i\sqrt{\epsilon}}
{\gamma}~
\int_{-\infty}^t\sin(t-\tau)
\sech^3(-\epsilon v\tau)\sinh(-\epsilon v\tau)
\cos\left(\sqrt{1-\epsilon^2}\tau\right) d\tau~, \nonumber\\ 
&=& -16\frac{\epsilon\epsilon_i\sqrt{\epsilon}}
{\gamma\sqrt{\lambda}}~
\cos(t)\int_{-\infty}^\infty \sin(\tau)
\sech^3(-\epsilon v\tau)\sinh(-\epsilon v\tau)
\cos\left(\sqrt{1-\epsilon^2}\tau\right) d\tau~, 
\nonumber\\
&\approx& -8\frac{\epsilon\epsilon_i\sqrt{\epsilon}}
{\gamma\sqrt{\lambda}}~
\cos(t)\int_{-\infty}^\infty \sin(2\tau)
\sech^3(-\epsilon v\tau)\sinh(-\epsilon v\tau) d\tau~,\nonumber\\
&=&0~.\label{eq-xshift}
\end{eqnarray}

We have used the same approximations as in Eq.~(\ref{eq-tshift}). For example, in the third and fourth lines we have set the frequency of the cosine to be $1$, which is correct in the leading order of $\epsilon$. Note an important difference from Eq.~(\ref{eq-tshift}). Instead of a $\cos^2()$ which is positive definite, we have an oscillation ($\sin()$) around zero. This leads to the integral being zero at the leading order\footnote{Note that without setting $\sqrt{1-\epsilon^2}\rightarrow1$, one could have gotten a nonzero answer from this integral. However, such answer is higher order in the $\epsilon$ expansion, which cannot be trusted in our approximation.}.

\begin{figure}[t!] 
   \centering
   \includegraphics[width=2.9in]{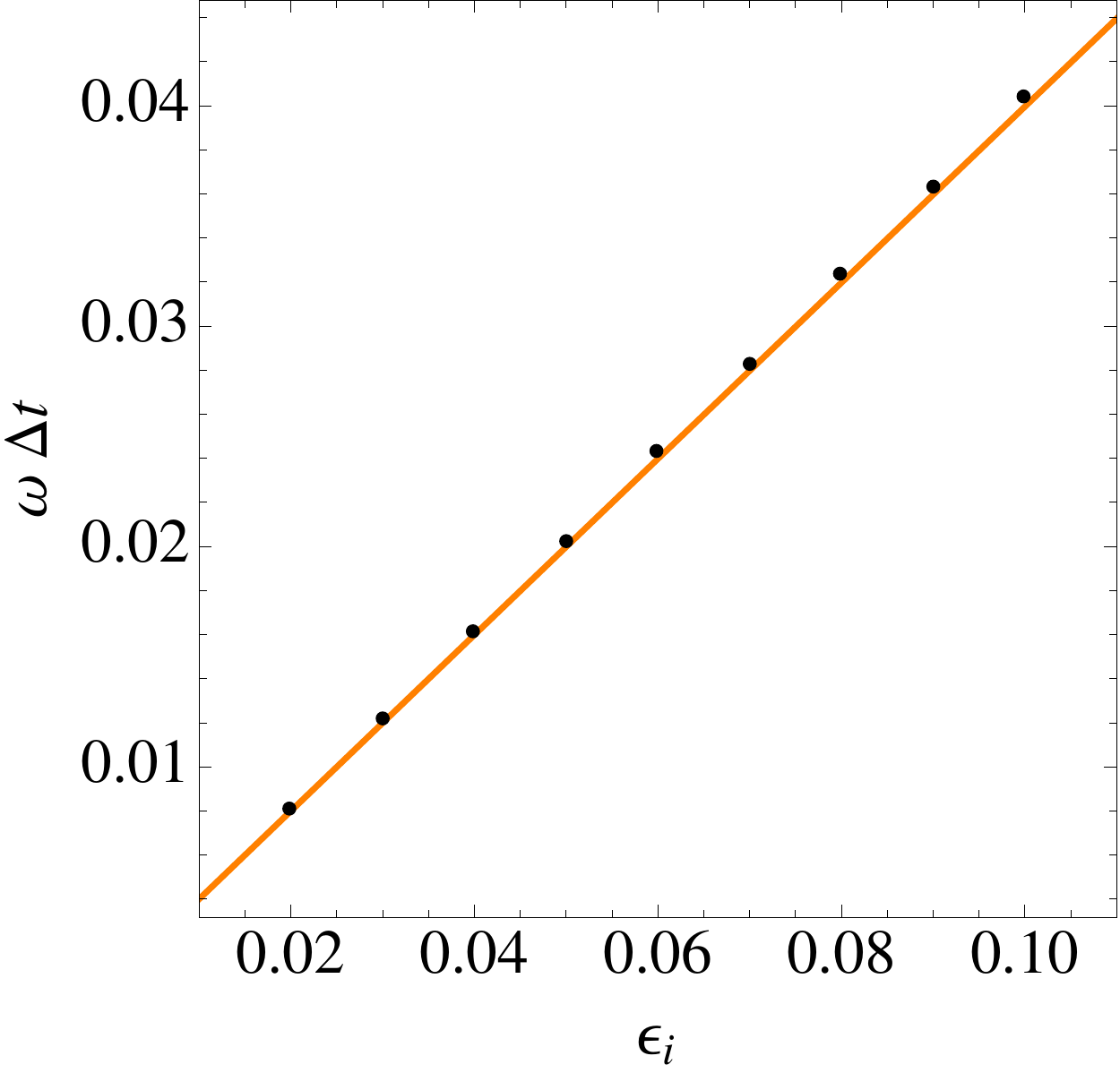} 
      \includegraphics[width=3.1in]{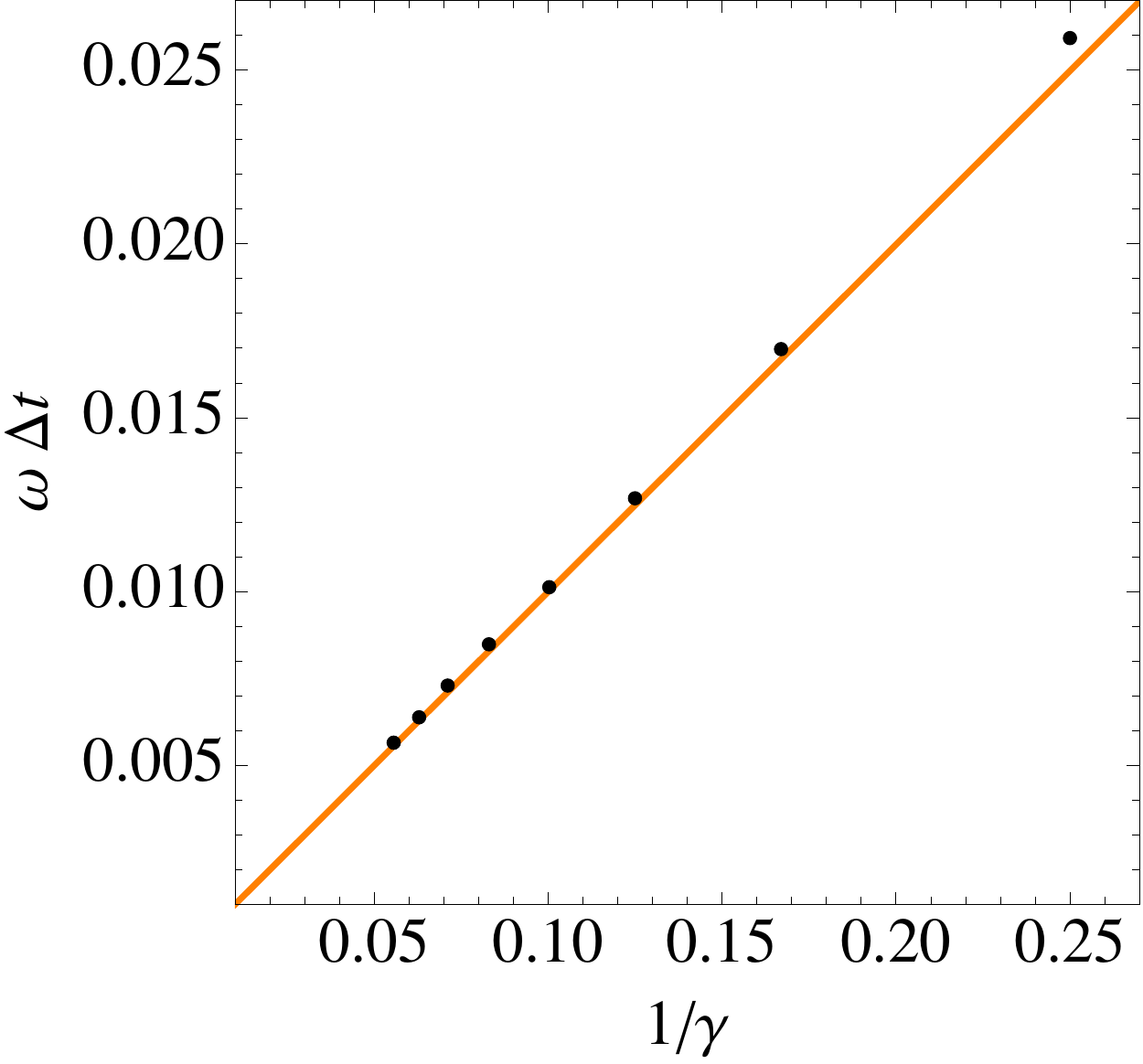}
   \caption{The dependence of the temporal phase shift on the amplitude of the incoming oscillon $\epsilon_i$ and the inverse boost factor $\gamma^{-1}$. The black dots indicate the results from numerical simulations, whereas the orange line is the expected leading order behavior in these small parameters based on our analytic calculation. In the right panel, the deviation is visible at $\gamma\sim4$, which shows that higher order terms in the expansion become important.}
   \label{fig:ParamDep}
\end{figure}

\section{Results}
\label{sec:results}
With the $f_n(t)$ solutions at hand, we are now ready to calculate the explicit expressions for $\Delta t$ and $\Delta x$.
Comparing Eq.~(\ref{eq-tshift}) with Eq.~(\ref{eq-ft}) and Eq. (\ref{eq-xshift}) with Eq. (\ref{eq-fx}), we get the leading order change in the position and temporal phase of the stationary oscillon after the collision:
\begin{eqnarray}
\omega \Delta t &=& \frac{\epsilon_i}{\gamma}
\left[4+\mathcal{O}(\epsilon)+\mathcal{O}(\epsilon_i)\right]~, \\
\epsilon\Delta x &=& \frac{\epsilon_i}{\gamma}
\left[0+\mathcal{O}(\epsilon)+\mathcal{O}(\epsilon_i)\right]~.
\end{eqnarray}
We have chosen to uphold the condition $\epsilon_i\gamma\gg\epsilon$ and kept the leading order terms in $\epsilon,\epsilon_i$ and $\gamma^{-1}$, a good approximation for ultra-relativistic, small amplitude collisions. Only the temporal phase shift gets a nonzero contribution at the leading order. As we explained in the end of Sec.\ref{sec:oscillons}, the lack of time dependence in $\Delta x$ (more explicitly a velocity term which would be linear in time) implies that the amplitude change $\Delta \epsilon$ is also zero. Thus, this collision is elastic at the leading order, same as the collision of kinks \cite{ALY13a}. These are the main results of our short paper.

To test our formalism and our analytical results, we carried out detailed 1+1 dimensional lattice simulations of the oscillon collisions.
For the temporal phase shift $\omega \Delta t$ (which is nonzero at leading order), we compared the results from numerical simulations (black dots) with the result from our analytic calculation for several values of $\gamma^{-1}$ and $\epsilon_i$.  Excellent agreement with our analytic answer in Eq.~(\ref{eq-tshift}) can be seen in Fig. \ref{fig:ParamDep}. The energy conservation in our simulations was better than 1 part in $10^4$. 

We have thus confirmed that the kinematic framework put forth in \cite{ALY13} can be applied to the case of time-dependent solitons. The excellent agreement between the numerical and analytical results is encouraging. We used small amplitude oscillons in 1+1 dimensions as our specific example of time dependent solitons. Understanding oscillon interactions is interesting in its own right, given their ubiquitous appearance in many physical scenarios from the end of inflation to condensed matter systems. It would be interesting to see if the agreement between our analytic results and simulations continues to hold beyond the 1+1 dimensional example considered here. A similar analysis should also be feasible for other time dependent solitons such as Q-balls\cite{Col85,Lee:1991ax}.

\section*{Acknowledgemements}
MA is supported by a Senior Kavli Fellowship at the University of Cambridge. ISY is supported by the research program of the Foundation for Fundamental Research on Matter (FOM), which is part of the Netherlands Organization for Scientific Research (NWO). 
\bibliography{all}
\bibliographystyle{utcaps}


\end{document}